# Reconfigurable Intelligent Surface-Empowered MIMO Systems[1]
*(Invited Paper)*

**Ertugrul Basar**
**Koc University, Department of Electrical and Electronics Engineering, Istanbul, Turkey**
ebasar@ku.edu.tr

**Abstract** – Reconfigurable intelligent surface (RIS)-empowered communication stands out as a solid candidate for future wireless networks due to its flexibility, ease of deployment, and attractive advantages to control the wireless propagation environment. In this perspective article, a brief overview is presented considering the application of reconfigurable intelligent surfaces for future multiple-input multiple-output (MIMO) systems. Potential future research directions are also highlighted.

5th Generation (5G) wireless networks, which are being deployed massively worldwide these days, provide higher flexibility and spectrum/energy efficiency compared to their earlier counterparts. However, 6th Generation (6G) wireless networks that are expected to roll out around 2030, will require more radical communication paradigms, especially at the physical layer of the communication stack. In this context, communication by means of reconfigurable intelligent surfaces (RISs) has been put forward as a potential candidate to satisfy the challenging requirements of future wireless networks in terms of energy efficiency and deployment cost [1]. In simple terms, an RIS can manipulate the propagation environment by performing unique functions such as anomalous reflection, over-the-air amplification, backscattering, refraction, absorption, etc. to boost the signal strength, alleviate the inter-channel interference and thus enhance the channel capacity gains. Specifically, the power consumption of a fully passive sub-6 GHz RIS, which can be controlled through USB, can be as low as a couple of watts. Due to their promising advantages, researchers considered the use of RIS technology in diverse communication scenarios. In this perspective article, we will provide a brief overview of RIS applications for emerging multiple-input multiple-output (MIMO) systems.

The generic model of an RIS-assisted MIMO system is given in Fig. 1 [2]-[4]. In a nutshell, in terms of MIMO interaction, an RIS can used for i) spatial multiplexing rank improvement, ii) joint active and passive beamforming, iii) RIS reflection modulation, and iv) virtual MIMO transmission. As shown in Fig. 1, depending on the availability as well as the condition of the direct channel between the source (S) and the destination (D), an RIS is used in different ways. For instance, when the channels between transmit and receive antennas have a strong correlation due to the existence of a strong line-of-sight communication link, the number of eigenchannels is limited for parallel data transmission. In this case, an RIS can be used for spatial multiplexing rank improvement by increasing the available degrees of freedom in the channel [4]. Similarly, in higher frequencies, the channel can be dominated by one or two strong paths and an RIS can be used to increase the spectral efficiency as well.

---

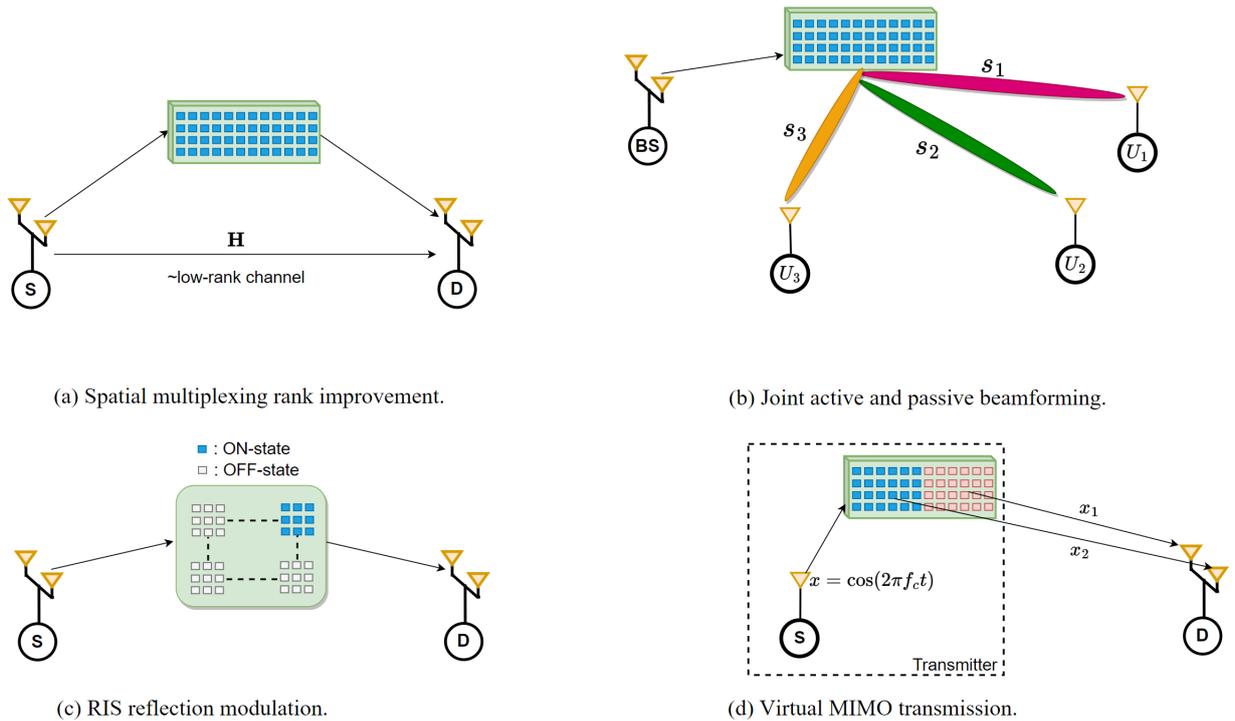

(a) Spatial multiplexing rank improvement.

(b) Joint active and passive beamforming.

(c) RIS reflection modulation.

(d) Virtual MIMO transmission.

Fig. 1: RIS-assisted MIMO communications with four different use-cases.

Alternatively, by jointly optimizing the transmit beamforming and RIS reflection phases, the received signal power, in return, the capacity can be boosted for a traditional MIMO system [2]. However, this joint optimization requires the availability of perfect channel state information (CSI) at the transmitter, and non-convex capacity maximization problems should be solved. At this point, alternating optimization-based methods can be considered to optimize the MIMO system capacity in the presence of an RIS. In [2], considering both frequency-flat and frequency-selective channels, the fundamental capacity limits of an RIS-aided point-to-point MIMO system are derived. Instead of complicated optimization problems, an efficient low-complexity algorithm, which is based on the cosine similarity theorem, can also be used to maximize the error performance of the target MIMO system by carefully adjusting the RIS phases [3].

The use of RISs can bring many opportunities for MIMO communication. Specifically, an RIS can be used to enhance the performance of the nulling and canceling-based suboptimal detection for a spatial multiplexing system [5]. In this way, spectral efficiency can be also boosted by conveying extra bits through the adjustment of the phases of the RIS elements. Another unique property of RISs is their operation as a virtual MIMO terminal [5]-[7]. In this case, an RIS illuminated by an unmodulated carrier can be configured to operate as a MIMO transmitter without any radio frequency (RF) chains. Specifically, in [5], a virtual space-time coding system utilizing Alamouti's coding is developed by carefully adjusting the RIS phases. In [6] and [7], RIS-based modulation, as well as multi-channel transmitter designs, are experimentally demonstrated, and virtual MIMO systems are created by using single and dual-polarized RIS architectures. Practical experiments from the literature have shown that an RIS can boost the received signal power up to 10 dB by carefully adjusting its phases and using directional antennas.

Beyond improving the signal power of the target receiver or the rank of a considered channel, RISs can also be used to suppress interference, e.g., inter-cell, inter-user, and inter-antenna interference by using its signal nulling capability [8]. For instance, in the case of a multi-antenna transmitter without any additional signal processing, an over-the-air beamforming solution is presented in [9] by eliminating inter-user interference. In [10], a synergy is created between index modulation (IM) and RISs by using an RIS to implement a receive IM scheme by focusing the reflected signal on one of the multiple antennas of a receiver.

Despite these attractive application areas for RIS-empowered MIMO systems, the inclusion of RISs in practical networks brings many challenging problems considering their passive nature. Particularly, wide-scale optimization of RISs should be performed to carefully position them in the network (i.e., base-station side or user-side deployments) and their real-time configuration without complicated signal processing and CSI acquisition. For instance, artificial intelligence-based solutions and predefined RIS codebooks can be used to determine the optimum RIS phase configurations to interact with existing MIMO systems, without utilizing full CSI knowledge. Metasurface-based RIS designs can be considered as well to obtain more flexibility in the system design despite their more complicated structure.

In conclusion, the past few years have witnessed significant interest from both academia and industry for the application of RISs for future wireless networks. In this context, the seamless integration of RISs to MIMO technologies, which lie at the heart of modern communication systems and standards, appears to be an interesting way to move forward. The growing interest from the industrial partners, for instance within the context of the ETSI RIS-Industry Specification Group, might be closely followed for future RIS deployments. Potential research directions for the following years include the development of simpler algorithms for the real-time adjustment of RIS-empowered MIMO systems, consideration of practical RIS setups with discrete phase elements, exploration of more advanced RIS architectures such as active RISs and simultaneous transmitting and receiving-RIS (STAR-RIS), and effective integration of RISs to massive and cell-free MIMO systems.

**Funding** – This work is supported by TUBITAK under Grand 120E401.